\documentclass[pre,twocolumn,amsmath,amssymb,nofootinbib,floatfix,superscriptaddress]{revtex4}

\usepackage{slashed} 
\usepackage{graphicx,bm,xcolor,bbold}
\usepackage{enumerate}
\usepackage{graphicx,bm}
\usepackage[normalem]{ulem}

\makeatletter
\def\graphicscale{\twocolumn@sw{0.3}{0.4}}
\def\graphicthreescale{\twocolumn@sw{0.3}{0.4}}

\begin{document}

\title{Conjecture on the lower bound of the length-scale critical
  exponent $\nu$ \\ at continuous phase transitions}

\author{Andrea Pelissetto}
\affiliation{Dipartimento di Fisica dell'Universit\`a di Roma
  ``La Sapienza" and INFN, Sezione di Roma I, I-00185 Roma, Italy}

\author{Ettore Vicari}
\affiliation{Dipartimento di Fisica dell'Universit\`a di Pisa,
  Largo Pontecorvo 3, I-56127 Pisa, Italy}

\date{\today}

\begin{abstract}
  A fundamental issue in the renormalization-group (RG) theory of
  critical phenomena concerns the allowed values of critical exponents
  that are consistent with the continuous nature of a phase
  transition.  Here we conjecture a lower bound for the length-scale
  exponent $\nu$, which should hold for the large class of continuous
  transitions associated with $d$-dimensional Landau-Ginzburg-Wilson
  (LGW) $\Phi^4$ theories with a multicomponent scalar field ${\bm
    \varphi}$ and a unique ${\bm \varphi}\cdot {\bm \varphi}$
  quadratic term (including some extensions with fermionic and gauge
  fields), describing many universality classes of critical phenomena.
  If $\Delta_\varphi=(d-2+\eta)/2$ is the dimension of the
  order-parameter field ${\bm \varphi}$, and
  $\Delta_\varepsilon=d-1/\nu$ is the RG dimension of the energy
  operator $\varepsilon$, which can be identified with $[{\bm
      \varphi}\cdot {\bm \varphi}]$ (the squared field with a proper
  subtraction of the mixing with the identity), we conjecture the
  inequality $\Delta_\varepsilon \ge 2 \Delta_\varphi$, which implies
  $\nu \ge (2-\eta)^{-1}$ and $\gamma = (2-\eta)\nu\ge 1$.  These
  inequalities are supported by general arguments for ferromagnetic
  lattice models, by $\epsilon$-expansion results for generic LGW
  $\Phi^4$ theories close to four dimensions, exact relations for
  two-dimensional minimal conformal field theories, and are consistent
  with all further known (numerical, perturbative, and exact) results
  for LGW $\Phi^4$ theories. In particular, since unitarity requires
  $\eta\ge 0$, the above inequality implies $\nu\ge 1/2$ for unitary
  theories.  This lower bound is more restrictive than $\nu > 1/d$,
  derived by noting that $\nu=1/d$ characterizes the singular
  finite-size behavior at first-order transitions.
\end{abstract}

\maketitle

\section{Introduction}
\label{intro}

The derivation of exact inequalities for the critical exponents
defined at continuous phase transitions is a crucial issue in the
theory of critical phenomena~\cite{Wilson-83, WK-74, Fisher-74,
  Wegner-76, PP-book, ZJ-book, PV-02}.  In this paper we conjecture an
exponent inequality that should hold at standard continuous
transitions admitting an effective description in terms of
$d$-dimensional Landau-Ginzburg-Wilson (LGW) $\Phi^4$
theories~\cite{ZJ-book,BLZ-74,BLZ-76,Aharony-76,PV-02,V-07}.

We consider critical behaviors characterized by the presence of a
single relevant symmetry-preserving perturbation, which may be
identified with the reduced temperature $t = (T-T_c)/T_c$ in many
cases. In LGW $\Phi^4$ theories with a multicomponent order-parameter
scalar field ${\bm \varphi}$, this scenario corresponds to
Hamiltonians with a single quadratic term $r \,{\bm \varphi}\cdot{\bm
  \varphi}$, where $r$ controls the approach to criticality in the
absence of external symmetry-breaking sources.  The length scale $\xi$
of the critical modes diverges as $\xi \sim |\bar r|^{-\nu}$, where
$\bar r = r-r_c$ is the deviation from the critical point. The
critical exponent $\nu$ depends on the universality class of the
continuous transition, which can be identified by a few global
properties of the critical modes, such as their spatial dimensions
$d$, their symmetry and symmetry-breaking pattern at the transition.

At continuous transitions the critical exponent $\nu$ must be larger
than $1/d$.  Indeed, in finite-size renormalization-group (RG)
analyses $\nu=1/d$ characterizes the emergent singular behavior of
first-order
transitions~\cite{NN-75,FB-82,PF-83,FP-85,CLB-86,Binder-87,BK-90,LK-91,BK-92,
  VRSB-93,PV-24}.  The lower bound $\nu > 1/d$ has often been used as
a criterion to distinguish continuous from first-order transitions in
numerical analyses. However, a systematic examination (as far as we
know, see below) of the results for the continuous transitions shows
that there are no analytic, (sufficiently robust) numerical or
experimental estimates of $\nu$ with $\nu<1/2$, suggesting that $\nu$
actually satisfies $\nu\gtrsim 1/2$. This inequality is significantly
stronger than $\nu>1/d$ for $d>2$ and, in particular, for $d=3$.

In this paper, we address this issue, and conjecture a lower bound of
the critical exponent $\nu$ for a large class of critical phenomena,
which is
\begin{equation}
  \nu \ge {1\over 2-\eta},
  \label{nubound}
\end{equation}  
where $\eta$ is the critical exponent associated with the power-law
decay $G({\bm x},{\bm y})\sim |{\bm x}-{\bm y}|^{-(d-2+\eta)}$ of the
critical two-point correlation function of the order-parameter field
(the exponent $\eta$ quantifies the deviation from the Gaussian
behavior).  Since $\eta\ge 0$ for critical behaviors associated with
unitary field theories, this bound explains the apparent absence of
continuous transitions with $\nu<1/2$. In the following we
often refer to the above lower bound for $\nu$ as the {\em $\nu$
  conjecture}.

The paper is organized as follows.  In Sec.~\ref{nuconj} we specify
the class of critical phenomena that we consider, we outline the main
ideas leading to the conjectured bound~(\ref{nubound}), and we discuss
related aspects and implications. In Sec.~\ref{boundgamma} we present
a general argument that supports the equivalent inequality $\gamma
=(2-\eta)\nu\ge 1$ for the susceptibility exponent $\gamma$ in generic
ferromagnetic lattice systems.  In Sec.~\ref{lgweps} we prove the
$\nu$ conjecture for LGW $\Phi^4$ theories close to four dimensions,
in the framework of the $\epsilon=4-d$ expansion.  In
Sec.~\ref{2dcont} we focus on two-dimensional (2D) continuous
transitions, and, in particular, we prove the $\nu$ conjecture for 2D
conformal minimal unitary models. In Sec.~\ref{lnlimit} we show that
the known analytical results for $d$-dimensional continuous
transitions in the limit of large number of field components agree
with the $\nu$ conjecture. In Sec.~\ref{3dcont} we show that all known
results for three-dimensional (3D) critical behaviors satisfy the
inequality~(\ref{nubound}). In Secs.~\ref{GNYmod} and \ref{AHmod} we
extend our discussion to models with fermionic fields and gauge
fields, such as the Gross-Neveu-Yukawa (GNY)~\cite{ZJ-book} and the
Abelian Higgs (AH)~\cite{ZJ-book,BPV-25-rev} model,
respectively. Finally, in Sec.~\ref{conclu} we draw our conclusions.
We also add a number of appendices reporting some details on the
results that are used to support the $\nu$ conjecture.\footnote{In
this paper we report results for a large number of critical phenomena
that support the conjecture (\ref{nubound}).  Whenever possible, we
will mainly refer to review papers and books, containing lists of
earlier references and original works.  The references that we report
are not meant to give credit to the many original papers that
addressed the different systems that we consider, which would require
a much larger number of references.  However, we generally mention the
references containing the most accurate results that we use in our
presentation.}

\section{The conjecture on the lower bound of the length-scale exponent}
\label{nuconj}

We consider the most general LGW
theory~\cite{ZJ-book,BLZ-74,BLZ-76,Aharony-76,PV-02,V-07} for an
$N$-component real field $\varphi_a$, $a=1,\ldots,N$, which has a
global symmetry group that forbids the presence of quadratic terms
proportional to $\varphi_a\varphi_b$ ($a\not= b$) and of cubic terms
in the Hamiltonian; for instance, this is the case if the model is
invariant under the parity ${\mathbb Z}_2$ transformations
$\varphi_a\to -\varphi_a$.  Thus, the Hamiltonian density takes the
form~\cite{ZJ-book,BLZ-74}
\begin{eqnarray}
  {\cal H} &=& \sum_a (\partial_\mu \varphi_{a})^2 + P({\bm \varphi}),
  \label{genH}\\
  P({\bm \varphi}) &=& \sum_a r_a \varphi_{a}^2 + \sum_{abcd} u_{abcd} \;
  \varphi_a\varphi_b\varphi_c\varphi_d, \nonumber
\end{eqnarray}
where $u_{abcd}$ is symmetric under any exchange of the indices (the
indices $a,b,c,d$ denote the field components, while the space
dependence is understood).  The number of independent parameters $r_a$
and $u_{abcd}$ depends on the symmetry group of the model. Standard
continuous transitions are described by LGW theories with a global
symmetry group that allows only one invariant quadratic term, given by
${\bm \varphi}\cdot {\bm \varphi} = \sum_a \varphi_a^2$.  We restrict
our analysis to this general class of LGW theories.\footnote{LGW
theories with more than one quadratic term describe multicritical
behaviors~\cite{PV-02,FN-74,NKF-74,CPV-03}. They are not considered in
this paper.} This implies $r_a = r$ for all $a$, and the trace
condition $\sum_a u_{aacd} = U \delta_{cd}$ for the quartic couplings
$u_{abcd}$~\cite{BLZ-74,ZJ-book}.  Criticality is approached by tuning
the single parameter $r$.  All field components become critical
simultaneously and the two-point function in the disordered phase is
diagonal:
\begin{equation}
\langle \varphi_a({\bm x}) \varphi_b({\bm y})\rangle =
\delta_{ab} G({\bm x}-{\bm y}).
\label{diagcondG}
\end{equation}
We will also consider extended LGW $\Phi^4$ theories, including
fermionic and gauge fields, such as the GNY~\cite{ZJ-book} and
AH~\cite{ZJ-book,BPV-25-rev} field theories.

In the LGW framework, the exponents $\eta$ and $\nu$ are related to
the RG dimensions $\Delta_\varphi$ and $\Delta_\varepsilon$ of the
field operator ${\bm \varphi}$ and of the {\em energy} operator
$\varepsilon \equiv [{\bm \varphi}^2]$, which is the square of the
field operator after subtracting the mixing with the
identity. Explicitly, we have
\begin{equation}
  \Delta_\varphi = {d-2+\eta\over 2},\quad
  \Delta_{\varepsilon} = d - y_r,\quad y_r={1\over \nu}
  \label{rgdimrel}
  \end{equation}
where $y_r$ is the RG dimension of the deviation $\bar r\equiv
r-r_c$~\cite{Wilson-83, WK-74, Fisher-74, Wegner-76, PP-book, ZJ-book,
  PV-02}.  One can easily verify that the conjectured inequality
(\ref{nubound}) is equivalent to the following inequality for the RG
dimensions $\Delta_{\varepsilon}$ and $\Delta_\varphi$:
\begin{eqnarray}
  \Sigma \equiv \Delta_{\varepsilon} - 2 \Delta_\varphi \ge 0.
  \label{conj1}
  \end{eqnarray}
Indeed, using Eq.~(\ref{rgdimrel}),
we can rewrite Eq.~(\ref{conj1}) as
\begin{eqnarray}
  \Sigma = 2 - \eta - {1\over \nu}  \ge 0,
\label{nusigma}
\end{eqnarray}
which is equivalent to Eq.~(\ref{nubound}).  The equality $\Sigma=0$
is satisfied in the Gaussian model, in which $\Delta_\varepsilon$ and
$\Delta_\varphi$ coincide with the {\em naive} dimensions:
$\Delta_\varepsilon=d-2$ and $\Delta_\varphi=(d-2)/2$.  Note that, if
the theory is unitary (corresponding to reflection positivity in
Euclidean space), then $\eta\ge
0$~\cite{PP-book,ID-book,PRV-19}. Therefore the inequality
(\ref{nubound}) implies $\nu\ge 1/2$ for unitary LGW theories, which
is stronger than $\nu>1/d$.

We may also consider the critical exponent $\gamma$ that parameterizes
the divergence $\chi \sim |\bar r|^{-\gamma}$ of the susceptibility
$\chi = \int d^dx\, G({\bm x})$. Since $\gamma$ is related to the
critical exponents $\nu$ and $\eta$ by the scaling
relation~\cite{Wilson-83, WK-74, Fisher-74, Wegner-76, PP-book,
  ZJ-book, PV-02}
\begin{equation}
\gamma = d- 2\Delta_\varphi=(2-\eta)\nu,
\label{gammascalr}
\end{equation}
we may equivalently write the inequality
(\ref{nubound}) as
\begin{equation}
  \gamma = (2-\eta)\,\nu \ge 1.
  \label{gammabound}
\end{equation}
We also recall that $\gamma$ must satisfy the upper bound
$\gamma<d\nu$ for consistency.

Inequality (\ref{conj1}) has a notable implication on the
operator-product expansion (OPE) of the operator product ${\bm
  \varphi}({\bm x}_1)\cdot {\bm \varphi}({\bm x}_2)$ when
$|{\bm x}_1-{\bm x}_2|\to 0$. Indeed, assuming
translation invariance, its OPE reads~\cite{ZJ-book,PP-book}
\begin{eqnarray}
  {\bm \varphi}({\bm x}_1)\cdot {\bm \varphi}({\bm x}_2) \approx
  F_I({\bm x})\;I + F_\varepsilon({\bm x}) \;
  [{\bm \varphi}^2]({\bm X}),
  \label{OPEexp}
\end{eqnarray}
where $I$ is the identity operator,
${\bm x} \equiv {\bm x}_1-{\bm x}_2$,
${\bm X} \equiv ({\bm x}_1+{\bm x}_2)/2$, 
\begin{eqnarray}
F_I({\bm x})= \langle {\bm \varphi}({\bm x}_1)\cdot {\bm
  \varphi}({\bm x}_2) \rangle,\qquad
F_\varepsilon({\bm x}) \propto |{\bm x}|^\Sigma.
\label{opevar}
\end{eqnarray}
Therefore $F_\varepsilon({\bm x})$ is nondiverging if the
inequality $\Sigma\ge 0$ holds. Moreover, conformal symmetry implies
that at the critical point~\cite{Polyakov-70,PP-book,ID-book,DMS-book}
\begin{eqnarray}
G_{\varphi\varphi\varepsilon}=
  \langle {\bm \varphi}({\bm x}_1) \cdot {\bm \varphi}({\bm x}_2) \;
\varepsilon ({\bm x}_3)\rangle \propto {|{\bm x}_{12}|^\Sigma
  \over |{\bm x}_{23}|^{\Delta_\varepsilon}
  |{\bm x}_{13}|^{\Delta_\varepsilon}} \label{cftthree2}
\end{eqnarray}
where ${\bm x}_{ij} \equiv {\bm x}_i-{\bm x}_j$.
Thus, $\Sigma\ge 0$ implies that
$G_{\varphi\varphi\varepsilon}$ does not diverge when ${\bm x}_{12}\to
0$ keeping ${\bm x}_{13}$ and ${\bm x}_{23}$ finite.

As we shall see, the conjecture (\ref{conj1}) and the equivalent inequalities
(\ref{nubound}) and (\ref{gammabound}) on $\nu$ and $\gamma$ are
supported by general arguments for lattice systems, and are consistent
with all known results for the critical exponents obtained
numerically, by using the $d=2,3,4-\epsilon$ expansions, the large-$N$
expansion, and conformal-field theory (CFT) in two dimensions.  The
$\nu$ conjecture should also apply to $d$-dimensional quantum
continuous transitions that can be related to  $(d+1)$-dimensional LGW
$\Phi^4$ field theories, by the quantum-to-classical
mapping~\cite{SGCS-97,Sachdev-book,RV-21}.

\section{The bound $\gamma\ge 1$ in lattice systems}
\label{boundgamma}

The inequality $\gamma\ge 1$ was proved for Ising models in any
dimension in Refs.~\cite{Baker-75,GJ-77,Lebowitz-74}.  In this section
we conjecture an analogous inequality for generic ferromagnetic
systems, proving it in some limiting cases (high-temperature limit and
low-temperature phase).

We consider an $N$-component real field $\phi_{\bm x}^a$ (where
$a=1,...,N$) defined on the sites ${\bm x}$ of a $d$-dimensional
lattice of volume $V$ (which we assume cubic for simplicity), and the
partition function\footnote{In principle, one might consider a more
general hopping term in the Hamiltonian $H$ of the form $\sum_{a,b}
\phi_{\bm x}^a M_{ab} \phi_{\bm y}^b$, where $M$ is a
positive-definite (to guarantee the model to be ferromagnetic)
symmetric matrix. The matrix $M$ can be written as $M = V^T M_D V$,
where $M_D = \hbox{diag }(\lambda_1,\ldots,\lambda_N)$ and $V$ is an
orthogonal matrix.  Then, the redefinition $\phi^a \to (\phi^a)' =
\sqrt{\lambda_a} (V\phi)^a$ allows us to recover the simple hopping
term appearing in Eq.~(\ref{lattpart}).}
\begin{equation}
  Z = \int \prod_{\bm x} d\mu({\bm \phi}_{\bm x})\;
  e^{-\beta H},\quad H=-\sum_{\langle {\bm x}{\bm y}\rangle} {\bm
      \phi}_{\bm x}\cdot {\bm \phi}_{\bm y}, 
    \label{lattpart}
\end{equation}
where the sum is over nearest-neighbor sites,
\begin{equation}
  d\mu({\bm \phi}_{\bm x}) = 
  e^{-P({\bm \phi}_{\bm x})} d{\bm \phi}_{\bm x},
 \label{measlat}
\end{equation}
and $P({\bm \phi}_{\bm x})$ is a local potential.  We assume that the
model undergoes a continuous transition at $\beta = \beta_c$, where
all components $\phi_{\bm x}^a$ become critical. The latter condition
is verified if the model is invariant under a transformation group $G$
such that the field transforms irreducibly under $G$. If we define
\begin{equation}
K_{ab} =  \int d\mu({\bm \phi})\, {\bm \phi}^a {\bm \phi}^b,
\end{equation}
then the matrix $K_{ab}$ can be rewritten as\footnote{Because of the
specific form of the Hamiltonian hopping term, the symmetry
transformations ${\bm \phi} \to T {\bm \phi}$ are implemented by
orthogonal matrices $T$. The invariance of $d\mu({\bm \phi})$ under
the symmetry group implies the relation $\sum_{cd} T_{ac} T_{bd}
K_{cd} = K_{ab}$ or, in matrix form, $T K T^T = K$. Schur's lemma then
implies that $K$ is proportional to the identity.}
\begin{equation}
K_{ab} = k \delta^{ab} \int d\mu({\bm \phi}) ,
\label{kconstant}
\end{equation}
with a positive constant $k$. As a consequence, the two-point function
of the order-parameter field $\phi_{\bm x}^a$ satisfies the relation
(\ref{diagcondG}).  We do not make any assumption on $P({\bm \phi})$,
which may depend on $\bm \phi$ in an arbitrary manner, which is only
constrained by the global symmetry of the model.  Therefore, the model
(\ref{lattpart}) is generic and valid for any symmetry group $G$.

For Ising models the proof \cite{Baker-75} of the bound $\gamma \ge 1$
exploits the following inequality proved by Lebowitz
\cite{Lebowitz-74}:
\begin{eqnarray}
  &&\langle \sigma_{\bm x} \sigma_{\bm y} \sigma_{\bm z} \sigma_{\bm t}\rangle 
      \le
  \label{inequality}\\
  &&\quad \langle \sigma_{\bm x} \sigma_{\bm y} \rangle \langle 
                   \sigma_{\bm z} \sigma_{\bm t}\rangle + 
     \langle \sigma_{\bm x} \sigma_{\bm z} \rangle \langle 
                   \sigma_{\bm y} \sigma_{\bm t}\rangle + 
     \langle \sigma_{\bm x} \sigma_{\bm t} \rangle \langle 
                   \sigma_{\bm y} \sigma_{\bm z}\rangle,
\nonumber
\end{eqnarray}
where $\bm x,\bm y,\bm z,\bm t$ label the lattice sites and
$\sigma_{\bm x} = \pm 1$.  The proof of this relation relies heavily
on the Ising nature of the spin variables, i.e., $\sigma_{\bm x}=\pm
1$, and it has been generalized to Ashkin-Teller models~\cite{Wu-82}
in Ref.~\cite{CS-00}.  If
\begin{equation}
\chi = {1\over V} \sum_{{\bm x}{\bm y}}
\langle \sigma_{\bm x}\sigma_{\bm y}\rangle,
\label{chiisi}
\end{equation}
then relation (\ref{inequality}) implies
\begin{eqnarray}
&&{\partial \chi\over \partial \beta} =
    {1\over V} \sum_{{\bm x}{\bm y}} \sum_{\langle {\bm z}{\bm t}\rangle} 
 \left[ 
 \langle \sigma_{\bm x} \sigma_{\bm y} \sigma_{\bm z} \sigma_{\bm t}\rangle -
 \langle \sigma_{\bm x} \sigma_{\bm y} \rangle 
 \langle \sigma_{\bm z} \sigma_{\bm t} \rangle
   \right]
\label{inequality-Ising}
\\
   &&\quad \le
    {1\over V} \sum_{{\bm x}{\bm y}} \sum_{\langle {\bm z}{\bm t}\rangle} 
 \left[ 
 \langle \sigma_{\bm x} \sigma_{\bm z} \rangle 
 \langle \sigma_{\bm y} \sigma_{\bm t} \rangle + 
 \langle \sigma_{\bm x} \sigma_{\bm t} \rangle 
 \langle \sigma_{\bm y} \sigma_{\bm z} \rangle \right] =  q \chi^2,
\nonumber 
\end{eqnarray}
where the first sum extends over all lattice sites, the second sum
over all lattice bonds, and we used Eq.~(\ref{inequality}) in the
second step.  This inequality holds for generic regular lattices, with
$q$ being the lattice coordination number. For cubic lattices
$q=2d$. Finally, one can straightforwardly prove 
that the rigorous
inequality (\ref{inequality-Ising}) implies $\gamma\ge 1$.
Indeed, if the transition is continuous, then
for $\beta < \beta_c$ close to the transition we have
\begin{equation}
\chi \approx a_0 (\beta_c - \beta)^{-\gamma},
\qquad a_0 > 0.
\label{chicrit}
\end{equation}
Then Eq.~(\ref{inequality-Ising}) implies
\begin{eqnarray}
   a_0 \gamma (\beta_c - \beta)^{-\gamma-1} \le {q a_0^2} 
       (\beta_c - \beta)^{-2\gamma},   
  \label{chicrit2}
\end{eqnarray}
and therefore
\begin{eqnarray}
  0 \le  (\beta_c - \beta)^{\gamma-1} \le {q a_0\over \gamma},
 \label{chicrit3} 
\end{eqnarray}
which is only possible if $\gamma \ge 1$. Note that also the reverse
statement holds: if $\gamma \ge 1$, close to the critical point the
inequality (\ref{inequality-Ising}) holds.  Using the
  inequality $\gamma \le (2-\eta)\nu$ proved for Ising systems by
  Fisher~\cite{Fisher-69}, we can rigorously derive the inequality
  $\nu\ge (2-\eta)^{-1}$ for Ising systems.

We conjecture that an inequality analogous to the one reported in
Eq.~(\ref{inequality-Ising}) also holds in the general models
defined in Eq.~(\ref{lattpart}). Defining the susceptibility as 
\begin{equation}
  \chi = {1\over V} \sum_{{\bm x},{\bm y}} \langle
    {\bm \phi}_{\bm x}\cdot {\bm \phi}_{\bm y}\rangle,
\end{equation}
we conjecture the existence of a positive constant
$C$ such that
\begin{equation} 
  W(\beta,C) \equiv \lim_{V\to\infty} \left( {\partial \chi\over \partial \beta} 
  - C \,\chi^2 \right) \le 0. 
\label{inequality-2}
\end{equation}
If this inequality holds, then the same argument outlined for the Ising
model [cf. Eqs.~(\ref{chicrit}), (\ref{chicrit2}) and
  (\ref{chicrit3})] implies $\gamma\ge 1$.

We do not have a conclusive proof of Eq.~(\ref{inequality-2}). Here,
we prove the inequality in the low-temperature ferromagnetic phase,
i.e., for any $\beta > \beta_c$, and in the high-temperature phase for
small values of $\beta$.  In the ferromagnetic phase, the inequality
is verified for any positive $C$.  As the system magnetizes, in the
large volume limit $\chi$ behaves as
\begin{equation}
\chi \approx A(\beta) V, \qquad A(\beta) = m(\beta)^2,
\label{chifer}
\end{equation}
where $m(\beta)$ is the magnetization. Thus,
$\partial\chi/\partial\beta$ scales as the volume $V$, while $\chi^2$
scales as $V^2$, so Eq.~(\ref{inequality-2}) is certainly true for
large values of $V$.  In the high-temperature phase, for $\beta$
small, the straightforward computation of the first few terms of the
small-$\beta$ expansion gives
\begin{equation}
{\chi \over k N} = 1 + 2d  k \beta + 2d (2d-1) k^2 \beta^2 + O(\beta^3),
\end{equation}
which implies
\begin{eqnarray}
{W\over k^2 N} = (2d - \widetilde{C}) + 4 d k (2d - \widetilde{C} - 1) \beta +
  O(\beta^2),\quad
  \label{Wsctext}  
\end{eqnarray}
where $k$ is the positive constant defined in Eq.~(\ref{kconstant})
and $\widetilde{C} = N C$.  If $\widetilde{C} \ge 2 d$, then
$W(\beta,C)$ is negative and decreasing, proving the inequality
(\ref{inequality-2}) for small $\beta$.  We have no proof for generic
values of $\beta < \beta_c$. However, since for $\widetilde{C}\ge 2 d$
the inequality holds and $W(\beta,C)$ decreases for small $\beta$, and
$W(\beta,C)=-\infty$ for $\beta>\beta_c$, it is plausible that the
inequality holds for any $\beta$.  Therefore, although a conclusive
proof is still missing, we believe that the above analysis strongly
supports the general bound $\gamma\ge 1$ for continuous transitions in
any ferromagnetic lattice system.

We finally note that the inequality (\ref{inequality-2}) can also be
used as a tool to try to understand the nature of the transition in
those cases in which fits of numerical data give estimates of $\nu$
that are not compatible with the bound $\nu \ge (2-\eta)^{-1}$.  If
one finds that the inequality (\ref{inequality-2}) is satisfied close
to the critical point in a region where infinite-volume results can be
reliably obtained, the inconsistent result for $\nu$ should be taken
as an indication that the transition is not continuous.

\section{LGW $\Phi^4$ theories close to four dimensions}
\label{lgweps}

In four dimensions, LGW $\Phi^4$ theories become Gaussian theories
with logarithmic corrections (triviality), see, e.g.,
Refs.~\cite{ZJ-book,ID-book}, and thus their critical behavior is
consistent with the $\nu$ conjecture.  The RG flow close to four
dimensions can be investigated in the framework of the $\epsilon=4-d$
expansion~\cite{WF-72,WK-74,Fisher-74,Wilson-83,BLZ-76,ZJ-book,WZ-74,VZ-06}.
In this approach one considers dimensional regularization and the
minimal-subtraction (MS) renormalization scheme.  The one-loop $\beta$
function of the MS renormalized couplings $g_{abcd}$, associated with
the Hamiltonian couplings $u_{abcd}$ in Eq.~(\ref{genH}), is given
by~\cite{BLZ-74,ZJ-book}
\begin{eqnarray}
&&\beta_{abcd}(g) = {\partial g_{abcd}\over \partial\mu} = -\epsilon
g_{abcd} \label{betaftext}\\ &&\;\; + {1\over 2}
(g_{abmn}\,g_{mncd}+g_{acmn}\,g_{mnbd}+g_{admn}\,g_{mnbc}),
\nonumber
\end{eqnarray}
where a sum is understood over repeated indices (and also
below). Moreover, the normalization of the couplings is chosen to
simplify the expressions, see, e.g., Ref.~\cite{ZJ-book}.  Note that,
like the quartic couplings $u_{abcd}$ of the $\Phi^4$ Hamiltonian
density, the MS renormalized couplings $g_{abcd}$ are completely
symmetric with respect to interchanges of the indices, and satisfy the
trace condition~\cite{BLZ-74,ZJ-book}
\begin{equation}
  g_{aacd} = T \delta_{cd}.
  \label{tracond}
\end{equation}
The fixed points (FPs) are obtained from the zeroes of the $\beta$
function.  The stable FP generally determines the critical behavior of
systems with the same symmetry as the LGW theory, as long as the FP is
related by RG transformations with an effective bare LGW Hamiltonian, 
which has the same symmetry-breaking pattern observed at the transition.  The
unstable FPs are instead associated with critical behaviors that can
only be observed by performing additional tunings of the
model parameters.

At any FP we have 
\begin{equation}
  g_{abcd}^* = \epsilon f_{abcd} + O(\epsilon^2),
  \qquad   T^* = t\,\epsilon + O(\epsilon^2).
\label{gstareps}
\end{equation}
Moreover, the FP equation $\beta_{aabb}(g^*) = 0$ implies
\begin{equation}
N t^2   - 2 N t + 2 f_{abcd} f_{abcd} = 0,
\end{equation}
where $N$ is the number of components of the field. This equation can
only be satisfied if
\begin{equation}
    0 \le t \le 2, \quad \qquad f_{abcd} f_{abcd} \le {N\over 2}.  
\label{relc}
\end{equation} 
Note that $t=0$ is only possible if $f_{abcd} = 0$.  Relation
(\ref{relc}) implies that $T^*$ is nonnegative close to four
dimensions.

The RG dimensions $\Delta_\varphi$ and $\Delta_{\varepsilon}$ of the
order-parameter field ${\bm \varphi}$ and of the composite energy-density
operator $\varepsilon= [{\bm \varphi}^2]$ can be obtained by
evaluating the corresponding RG functions at the FP.
The RG functions associated with $\eta$ and $y_r=1/\nu$ are given
by~\cite{BLZ-74,ZJ-book}
\begin{eqnarray}
    && \eta(g) = {1\over 24N} g_{abcd} \,g_{abcd} +
  O(g^3), \label{etapert}\\
    && y_r(g) = 2 - {1\over 2 N} g_{aabb} + O(g^2) =
  2 - {1\over 2} T + O(g^2), \nonumber 
\end{eqnarray}
where we used the trace condition (\ref{tracond}). 
For any FP, we obtain 
\begin{equation}
  y_r = {1\over\nu} = 2 - {t\over 2} \,\epsilon + O(\epsilon^2).
  \label{nugfp}
\end{equation}
Since $\eta=O(\epsilon^2)$, we have 
\begin{eqnarray}
  \Sigma  = 2- \eta- y_r = {t\over 2} \,\epsilon + O(\epsilon^2)
  \ge 0,
  \label{kappaeps}
  \end{eqnarray}
proving the $\nu$ conjecture close to four dimensions for any,
stable or unstable, FP of LGW theories.

\section{Two-dimensional continuous transitions}
\label{2dcont}

The $\nu$ conjecture can be also proved for the 2D conformal minimal
unitary models~\cite{ID-book,DMS-book,BPZ-84,FQS-84} with central
charge
\begin{equation}
  c=1-{6\over m(m+1)},\qquad  m\ge 3,
  \label{centralch}
  \end{equation}
  when the energy operator
$\varepsilon$ can be associated with one of the primary operators with
dimension $2 h_{m-1,m-1}=2 h_{1,2}$ and $2 h_{2,1}$.\footnote{Minimal
models have a finite number of primary operators with scaling
dimensions parameterized by~\cite{BPZ-84,FQS-84}
\begin{equation}
h_{p,q} = h_{m-p,m+1-q} = {[(m+1)^2 p - mq]^2 - 1 \over 4 m (m+1)}.
\nonumber
\end{equation}
Because of the symmetry, one needs only to consider $1\le p\le m-1$
and $1\le q \le p$.}  This is the case of the Ising and three-state
Potts model, corresponding to $m=3$ and $m=5$ respectively, for which
$\Delta_\varepsilon=2\,h_{2,1}$, and of the tricritical Ising and
three-state Potts model, corresponding to $m=4$ and $m=6$
respectively, for which $\Delta_\varepsilon = 2\, h_{1,2}$.  For this
class of models one can derive an exact relation between
$\Delta_\varepsilon$ and $\Delta_\varphi$.

To prove the $\nu$ conjecture for minimal unitary models, we
consider the three-point correlation function
\begin{equation}
  G_{\varepsilon\varphi\varphi} \equiv \langle \varepsilon(z_1,\bar{z}_1)
  \varphi(z_2,\bar{z}_2) \varphi(z_3,\bar{z}_3)\rangle
  \label{a123def}
  \end{equation}
  of the energy operator $\epsilon$ and of two field operators
  $\varphi$ (we use complex spatial variables $z$). At the critical
  point, because of conformal invariance, it takes the
  form~\cite{Polyakov-70,PP-book}
\begin{equation}
  G_{\varepsilon\varphi\varphi} \propto 
    {z_{23}^{\Delta_\varepsilon/2-\Delta_{\varphi}} \over
    z_{12}^{\Delta_\varepsilon/2}z_{13}^{\Delta_\varepsilon/2}}\times
  ({\rm c.c.}),
\label{threeeq}
\end{equation}
where $z_{ij}=z_i-z_j$. When the energy operator $\varepsilon$ is a
primary operator with weight $h_{1,2}$ or
$h_{2,1}$~\cite{BPZ-84,ID-book}, the three-point function
$G_{\varepsilon\varphi\varphi}$ satisfies the differential equation
\begin{equation}
  \left[ {3\over 2(\Delta_\varepsilon+1)} \partial_{z_1}^2 -
    \sum_{k=2}^3 \left( {\Delta_{\varphi}\over 2(z_1-z_k)^2} +
        {\partial_{z_k}\over z_1-z_k}\right)\right]
  G_{\varepsilon\varphi\varphi} = 0.
  \label{diffeq}
\end{equation}
Substituting $G_{\varepsilon\varphi\varphi}$ given in Eq.~(\ref{threeeq}) into
Eq.~(\ref{diffeq}), one can easily derive the consistency condition
\begin{equation}
  2 \Delta_\varphi = {2 - \Delta_\varepsilon\over 2 (1 +
    \Delta_\varepsilon)} \Delta_\varepsilon .
  \label{conscond2dCFT}
  \end{equation}
This implies
\begin{equation}
  \Sigma = \Delta_\varepsilon-2 \Delta_\varphi = {3
    \Delta_\varepsilon^2 \over 2 (1 + \Delta_\varepsilon) } \ge 0,
\label{Sigma-minimal}
\end{equation}
which proves the inequality (\ref{conj1}) for all minimal
models.\footnote{Relation (\ref{conscond2dCFT}) also applies to the
four-state Potts model for which $\nu=2/3$ and $\eta=1/4$ (apart from
logarithmic corrections), which may be identified as the unitary model
obtained by the $m\to\infty$ limit of Eq.~(\ref{centralch}), with
central charge $c=1$.}

One can also check that the $\nu$ conjecture holds for other known 2D
critical phenomena, such as the Berezinskii-Kosterlitz-Thouless
transition~\cite{KT-73,Berezinskii-70,Kosterlitz-74,JKKN-77}
($\Sigma=7/4$), the O($N$) $\sigma$ models with $N\ge 3$ ($\Sigma=2$),
the O($N$) $\sigma$ models with $N\in
[-2,2]$~\cite{CH-80,Nienhuis-84,DF-84,CPRV-96} ($\Sigma$ increases
from $\Sigma=0$ for $N=-2$ to $\Sigma = 7/4$ for $N=2$), and the
transition line of the 2D Ashkin-Teller
model~\cite{Baxter-book,Baxter-71,Baxter-72,WL-74,DR-79} where
$\Sigma$ varies in the range $[1/4,\,5/4]$.  Some details are reported
in App.~\ref{2dex}.

\section{Continuous transitions in the large-$N$ limit}
\label{lnlimit}

The $\nu$ conjecture in confirmed in the large-$N$ limit for some
classes of LGW $\Phi^4$ theories, for which analytical results have
been obtained for any spatial dimension $d\in [2,4]$.  Some details are
reported in App.~\ref{lnvarious}.  For the O($N$) vector models,
see. e.g., Refs.~\cite{Ma-book,MZ-03}, $\Sigma$ is given by
\begin{eqnarray}
&&\Sigma =  4-d - Y(d)/N + O(N^{-2}),          
  \label{lnWondtext}\\
&& Y(d) = {4 (2d^2-7d+8)\Gamma(d-2)\over d\,\Gamma(d/2)
    \Gamma(2-d/2) \Gamma(d/2-1)^2}, \nonumber
\end{eqnarray}
which implies $\Sigma\ge 0$ for $d\in[2,4]$.  Analogous computations
have been performed in the $d$-dimensional
O($M$)$\otimes$O($N$)-symmetric LGW theory with 
potential~\cite{Kawamura-91}
\begin{equation}
  P = r \sum_{ai} \Phi_{ai}^2 + u \Bigl( \sum_{ai} \Phi_{ai}^2 \Bigr)^2 +
  v \sum_{aibj} \Phi_{ai}\Phi_{bi} \Phi_{aj}\Phi_{bj},
\label{omnpot}
\end{equation}
where $\Phi_{ai}$ is an $M\times N$ matrix field.
For $N\to\infty$ keeping 
$M$ fixed, one obtains \cite{PRV-01b} 
\begin{equation}
\Sigma = 4-d - Y(d){(M+1) \over 2N} + O(N^{-2})
\label{lnWOMNctext}
\end{equation}
at the stable {\em chiral} FP, and
{\begin{equation}
  \Sigma = Y(d){(M+2)(M-1)\over 2MN} + O(N^{-2})
\label{lnWOMNctext2}
\end{equation}
at the unstable {\em antichiral} FP.  For both FPs, $\Sigma\ge
0$.

\section{Three-dimensional continuous transitions}
\label{3dcont}
      
We now show that the $\nu$ conjecture is verified by all
known results for 3D continuous transitions, see App.~\ref{3dex} for
more details.

We begin with the 3D O($N$) vector 
model~\cite{ZJ-book,PV-02,PRV-19,GZ-98,CPRV-02,Hasenbusch-10,
  KPSV-16,KP-17,FXL-18,
  Hasenbusch-21,CHPV-06,Hasenbusch-19,CLLPSSV-20,Hasenbusch-25,
  CHPRV-02,HV-11,Hasenbusch-20,Chester-etal-21,Clisby-10,Clisby-17,
  Hasenbusch-22}.  Using the presently most accurate estimates of the 
critical exponents, we
obtain $\Sigma=0.37565(1)$ for the Ising universality class ($N=1$),
$\Sigma=0.47312(5)$ for the XY universality class ($N=2$),
$\Sigma=0.5570(2)$ for Heisenberg universality class ($N=3$), $\Sigma
\approx 1-40/(3 \pi^2 N)$ for large $N$, and $\Sigma = 0.26711(4)$ for
self-avoiding walks ($N\to 0$).

The $\nu$ conjecture also holds
for transitions associated with 3D LGW theories with cubic anisotropy
(relevant for anisotropic magnets and randomly dilute spin models),
for which the scalar potential reads~\cite{Aharony-76}
\begin{eqnarray}
P({\bm \varphi})= r \,{\bm \varphi}^2  + u \,({\bm \varphi}^2)^2 + v
\sum_{a=1}^N \varphi_a^4 . \label{cubicpot}
\end{eqnarray}
For $N=2$ the stable FP has an enlarged O(2) symmetry, while for $N\ge
3$ the stable FP is anisotropic, see App.~\ref{3dex}.  The available
results, see, e.g., Refs.~\cite{Aharony-74,HL-74,
  Aharony-76,PV-02,CPV-00,HV-11,Chester-etal-21,Hasenbusch-23,
  Hasenbusch-24,Fisher-68,Aharony-73,Emery-75,GL-76,PV-00,HPPV-07},
show that $\Sigma>0$: we find $\Sigma = 0.5556(5)$ for $N=3$, $\Sigma
= 0.574(1)$ for $N=4$, $\Sigma = 0.54981(1)$ in the large-$N$ limit,
and $\Sigma = 0.500(4)$ in the $N\to 0$ limit (corresponding to the
randomly diluted Ising universality class~\cite{GL-76,PV-02}).

We also consider
transitions that are expected to be described by
the 3D O(2)$\otimes$O($N$) LGW theories~\cite{Kawamura-98,PV-02,V-07},
cf. Eq.~(\ref{omnpot}).  The existence of the chiral
O(2)$\otimes$O(2), chiral O(2)$\otimes$O(3) and $^3$He superfluid
universality classes was pointed out in
Refs.~\cite{Kawamura-98,PRV-01a,PV-02,CPPV-04,DPV-04} (see also
Refs.~\cite{DMT-04,NT-14,RRSR-25} for some opposite results).
Using their results, one finds $\Sigma>0$ at all RG FPs.
More complex LGW theories have been discussed in 
Refs.~\cite{V-07,DPV-06,PSV-08,PV-13}. The corresponding estimates
of the critical exponents satisfy $\Sigma>0$ as well.

Finally, we mention that the inequality $\Sigma\ge 0$ also holds for the
percolation transition---a particular limit of the
$\Phi^3$ field theory---in any dimension $d$ satisfying $3 \le d \le
6$~\cite{BFMMPR-99,XWLD-14,AMAH-90,Gracey-15,LW-19,Ziff-20,SVZ-20}.

\section{Gross-Neveu-Yukawa  models}
\label{GNYmod}

We can also show that the $\nu$ inequality (\ref{nubound}) holds in
GNY theories in which a real scalar field $\varphi$ is coupled with
$N_f$ fermionic fields $\psi_f({\bm x})$ where $f=1,\ldots, N_f$,
with Hamiltonian density~\cite{ZJ-book,MZ-03,BB-87}
\begin{equation}
{\cal H}=(\partial_\mu \varphi)^2 + r \,\varphi^2 + u \,\varphi^4
-\sum_{f=1}^{N_f} \bar{\psi}_f (\slashed{\partial} + m +
g \varphi)\psi_f.
     \label{lagrangianGNY}
\end{equation}
Each {\em flavor} component $\psi_f$ is a four-dimensional spinor, so
the total number $N$ of fermionic components is given by $N = 4 N_f$,
and the matrices $\gamma_\mu$ are the usual Euclidean 4$\times$4
matrices used in 4 dimensions \cite{BB-87}.  We again identify the
energy operator as $\varepsilon=[\varphi^2]$.

Therefore, in these models $\Sigma=\Delta_\varepsilon-2\Delta_\varphi
= 2 - \eta_\varphi - y_r$.  The known leading terms of the large-$N_f$
expansion of the exponents $y_r=1/\nu$ and $\eta_\varphi$ for the
$d$-dimensional GNY
model~\cite{MZ-03,Gracey-93,Gracey-94,EIKLPS-22,BFPV-23,CL-13,
  CS-07,KSS-98} are reported in App.~\ref{lnGN3}. They lead to
\begin{eqnarray}
  \Sigma   \approx {Z(d)\over N_f},\;\; Z(d) = {2
    (d-1) \Gamma(d-2)\over \Gamma(d/2) \Gamma(2-d/2) \Gamma(d/2-1)^2},
  \label{sigmagnyln}
\end{eqnarray}
which is positive since $Z(d)\ge 0$ for any $d\in [2,4]$.  Large-$N_f$
results are also available for the chiral XY and Heisenberg GNY
models~\cite{Gracey-21,Gracey-18}: they imply $\Sigma\ge 0$ as well.
Moreover, close to four dimensions, the $\epsilon$-expansion results
reported in Refs.~\cite{ZMMHS-17,INMS-18} (see App.~\ref{lnGN3}),
imply $\Sigma = B(N_f) \epsilon$ with $B(N_f)>0$, for all GNY
formulations.

It is also worth mentioning the case of the extended GNY models
coupled to the U(1) gauge field $A_\mu$, whose Hamiltonian density
reads
\begin{eqnarray}
{\cal H}&=&(\partial_\mu \varphi)^2 + r \,\varphi^2 + u \,\varphi^4
+ {1\over 4 e^2} F_{\mu\nu}^2  \label{lagrangianQEDGN}\\
&&-\sum_{f=1}^{N_f} \bar{\psi}_f
 (\slashed{D} + m + g\varphi) \psi_f,
\nonumber
\end{eqnarray}
where $F_{\mu\nu}\equiv \partial_\mu A_\nu - \partial_\nu A_\mu$,
$D_\mu = \partial_\mu + i A_\mu$ is the covariant derivative, and
$\psi_f$ are four-component Dirac fields.  Again the relation
$\Sigma=2-\eta_\varphi-y_r>0$ is supported by the analytical analyses
of the RG flow of this theory, within the $\epsilon$-expansion and
large-$N_f$ frameworks, see, e.g.,
Refs.~\cite{JH-17,ZMBM-18,BRM-19}. Indeed, using these results, we
obtain $\Sigma = B(N_f)\epsilon + O(\epsilon^2)$ with $B(N_f)>0$, and
the 3D large-$N_f$ behavior
\begin{equation}
  \Sigma = {8\over \pi^2 N_f} + O(N_f^{-2}).
  \label{siggnyqed3}
\end{equation}  

\section{Abelian Higgs gauge models}
\label{AHmod}

The $\nu$ conjecture also holds at the charged continuous transitions
of lattice AH gauge models (with some specifications related to the
presence of gauge invariance), where both scalar and U(1) gauge modes
become critical~\cite{BPV-25-rev}. Their critical behavior can be
described in terms of an effective AH field
theory~\cite{HLM-74,ZJ-book,BPV-25-rev}
\begin{eqnarray}
{\cal H} = \frac{1}{4 e^2} \,F_{\mu\nu}^2+ |D_\mu{\bm\phi}|^2 + r\,
\bar{\bm \phi}\cdot{\bm \phi} + u \,(\bar{\bm \phi}\cdot{\bm
  \phi})^2,
\label{AHtext}
\end{eqnarray}
in which a complex $N$-component scalar field ${\bm \phi}$ is coupled
with a U(1) gauge field $A_\mu$, $F_{\mu\nu}=\partial_\mu A_\nu
- \partial_\nu A_\mu$ and $D_\mu = \partial_\mu + i A_\mu$.

As for LGW $\Phi^4$ theories, the energy-density operator can be
identifed with $\varepsilon \sim [\bar{\bm \phi}\cdot {\bm \phi}]$,
which is well defined as it is gauge invariant.  On the other hand, it
is less obvious what can be identified as order parameter. In LGW
models without gauge invariance one considers the field ${\bm
  \phi}$. However, in the absence of a gauge fixing, in AH models the
field $\bm \phi$ is not gauge invariant, and therefore its correlation
functions are ultralocal, forbidding the definition of a corresponding
RG dimension $\Delta_\phi$. In this case, the  appropriate
order parameter is a nonlocal gauge-invariant field, which corresponds
to the Lagrangian field in the hard Lorenz gauge $\partial_\mu A_\mu =
0$~\cite{BPV-25-rev,BPV-24-ncAH}. In the perturbative setting, this
implies that $\Delta_\phi=(d-2+\eta_\phi)/2$ is the RG dimension of
$\bm \phi$ for $\zeta = 0$, where $\zeta$ is the Lorenz-gauge
parameter. Some details on the RG flow of AH field theories are
reported in App.~\ref{lnAH}.

In AH models charged continuous transitions occur only for $N >
N_*(d)$. Close to four dimensions, $N_*$ is large: $N>N_*(4)=
90+24\sqrt{15} \approx 183$.  Using the perturbative expansions
reported in Refs.~\cite{HLM-74,IZMHS-19}, see also App.~\ref{lnAH}, we
find $\Sigma \approx A(N)\,\epsilon$ with $A(N)>0$ for $N>N_*(4)$.
For 3D AH models the Higgs transition can be continuous for $N>N_*(3)$
with $N_*(3)\approx 7$~\cite{BPV-22,IZMHS-19,SJCSJY-25}. Using the
numerical results of Ref.~\cite{BPV-22} and the large-$N$ results of
Refs.~\cite{HLM-74,IKK-96,KS-08,BPV-25-rev}, $\Sigma > 0$ is obtained
in all cases.  For $N<N_*(3)$ the Higgs transitions are expected to be
of first order, with the notable exception of the one-component AH
model (which is related to superconductivity, see, e.g.,
Refs.~\cite{HLM-74,DH-81,HT-96,Herbut-book}). In this case the
transition belongs to the inverted XY universality
class~\cite{BPV-25-rev,NRR-03,BPV-24-decQ2}, related to the standard
XY universality class by duality. The exponent $\nu$ is the same as in
the XY
model~\cite{PV-02,CHPV-06,Hasenbusch-19,CLLPSSV-20,Hasenbusch-25},
$\nu\approx 0.6717$. On the other hand, since the order parameter in
the AH model (a nonlocal gauge-invariant operator) differs from the
order parameter in the XY model (duality only relates the energy
sectors), $\eta_\phi=-0.74(4)$ in the AH model~\cite{BPV-25-rev},
which differs from $\eta\approx 0.03$ for the XY model~\cite{PV-02}.
Using these estimates we obtain $\Sigma=1.25(4)$, confirming the $\nu$
conjecture.

\section{Conclusions}
\label{conclu}

A fundamental issue in the RG theory of critical phenomena is the
allowed range of values for the length-scale critical exponent $\nu$.
We conjecture the inequality $\nu\ge (2-\eta)^{-1}$ or, equivalently
$\gamma = (2-\eta)\nu\ge 1$. This should hold for a large class of
continuous transitions that admit an effective LGW $\Phi^4$
description in terms of a multicomponent scalar field ${\bm \varphi}$
and a unique quadratic term ${\bm \varphi}^2$.  These LGW theories are
obtained when the global symmetry group is sufficiently large to admit
only the invariant quadratic term ${\bm \varphi}\cdot {\bm \varphi}$
in the Hamiltonian density.  We also consider some extensions with
fermionic and gauge fields, such as the GNY and AH models. LGW
theories with a single quadratic term describe a large class of
critical phenomena characterized by the presence of a single relevant
symmetry-preserving perturbation, which may be identified with the
reduced temperature in many cases.
We do not consider LGW theories with multiple quadratic terms, which
describe more general multicritical
behaviors~\cite{PV-02,FN-74,NKF-74}, for which it is not clear how to
derive lower bounds analogous to those related to the $\nu$
conjecture.

In the LGW framework, the conjectured lower bound $\nu\ge
(2-\eta)^{-1}$ is equivalent to an inequality for the RG dimensions of
the energy operator $\varepsilon = [{\bm \varphi}^2]$ (the square of
the field operator after subtracting the mixing with the identity) and
of the order-parameter field ${\bm \varphi}$.  Indeed, it is
equivalent to the inequality $\Delta_\varepsilon \ge 2 \Delta_\varphi$
between the RG dimension $\Delta_\varepsilon=d-1/\nu$ of
$\varepsilon$ and the RG dimension $\Delta_\varphi=(d-2+\eta)/2$ of
${\bm \varphi}$.

We have shown that these inequalities are supported by general
arguments for lattice models, exact results for the RG flow in generic
LGW $\Phi^4$ theories close to four dimensions, exact relations for 2D
minimal CFTs, and are consistent with all known (numerical,
perturbative, and exact) results in any dimension.  It is important to
remark that, since unitarity requires $\eta\ge 0$, the inequality
$\nu\ge (2-\eta)^{-1}$ implies $\nu\ge 1/2$ for critical behaviors
described by unitary theories.  For $d>2$ this lower bound is
significantly more restrictive than the lower bound $\nu > 1/d$,
obtained by noting that $\nu=1/d$ characterizes the singular
finite-size behavior at first-order
transitions~\cite{NN-75,FB-82,PF-83,FP-85,CLB-86,Binder-87,BK-90,LK-91,BK-92,
  VRSB-93,PV-24}.

We finally remark that the lower bound $\nu\ge (2-\eta)^{-1}$ explains
the phenomenological fact that no continuous transitions with $\nu <
1/2$ have been found, both in theoretical and experimental studies.
Moreover, it may be particular useful for the identification of the
nature of the transition in numerical analyses of 3D systems.  Indeed,
if it holds, then finite-size behaviors with effective exponents
$\nu<1/2$ should be interpreted as crossover behaviors toward a
first-order transition, without the need of verifying that the data
scale with exponent $\nu\approx 1/d$, as it should occur
asymptotically.

Of course, additional studies are needed to assess the generality of
the inequalities conjectured in this paper, and, if violations are
found, to identify the reasons of its failure.

\acknowledgments

We thank Claudio Bonati, Mihail Mintchev, and Omar Zanusso for
interesting and useful discussions.

\appendix

\section{Results for two-dimensional  transitions}
\label{2dex}

In this section we report some details on 2D systems, extending the
discussion of Sec.~\ref{2dcont}.  In Table~\ref{2Dexp} we report the
critical exponents $\nu$, $\eta$, $\gamma$, and the difference
$\Sigma=2-\eta-y_r$, for several 2D models. We consider the Ising
model (whose critical behavior has central charge is $c=1/2$), the
$q=3$ and $q=4$ Potts model ($c=4/5$ and $c=1$ respectively), the XY
model, which undergoes a Berezinskii-Kosterlitz-Thouless (BKT)
transition~\cite{KT-73,Berezinskii-70,Kosterlitz-74,JKKN-77}, the
self-avoiding walk (SAW) model, the tricritical Ising and $q=3$ Potts
models ($c=7/10$ and $c=6/7$ respectively), and percolation, which
corresponds to the $q\to 1$ limit of the Potts model.  See, e.g.,
Refs.~\cite{ZJ-book,ID-book,PV-02,Wu-82,BPZ-84,FQS-84}.  They confirm
the inequality $\Sigma = 2-\eta-y_r \ge 0$.

The O($N$) models for $N\ge 3$ do not undergo finite-temperature
transitions in two dimensions. A transition occurs for
$d=2+\epsilon_2$ with $\epsilon_2>0$, with exponents \cite{ZJ-book}
$\nu = 1/\epsilon_2$ and $\eta =\epsilon_2/(N-2)$ (they formally
converge to $\infty$ and $0$, as reported in Table~\ref{2Dexp}, for
$\epsilon_2\to 0$).  It follows
\begin{equation}
\Sigma = 2 - {N-1\over N-2} \epsilon_2 + O(\epsilon_2^2) > 0.
\label{kadm2exp}
\end{equation}

\begin{table}
  \caption{Critical exponents $\nu$, $\eta$ and $\gamma$ for some 2D
    critical behaviors, and corresponding difference
    $\Sigma=\Delta_\varepsilon-2\Delta_\varphi$.  }
\label{2Dexp}
\begin{tabular}{ccccc} \hline\hline
  2D Model & $\quad \nu \quad$ & $\quad\eta\quad$
  & $\quad \gamma \quad$ & $\quad \Sigma \quad$ \\\hline

  Ising       & 1 & 1/4 & 7/4 & 3/4 \\

  $q=3$ Potts & 5/6 & 4/15 & 13/9 & 8/15 \\

  Tricritical Ising & 5/9 & 3/20 & 37/36 & 1/20\\

  Tricritical $q=3$ Potts & 7/12 & 4/21 & 19/18 & 2/21 \\
 
  $q=4$ Potts & 2/3 & 1/4 & 7/6 & 1/4 \\

  XY [BKT, O(2) $\sigma$]  &  $\infty$  & 1/4 & $\infty$ & 7/4 \\

  O($N\ge 3$) $\sigma$ &  $\infty$  & 0 & $\infty$ & 2\\

  SAW  [O($N\to 0$) $\sigma$]  &  3/4 & 5/24 & 43/32 & 11/24 \\ 

  Percolation ($q\to 1$ Potts) & 4/3 & 5/24 & 43/18 & 25/24 \\
\hline\hline
\end{tabular}
\end{table}

The $\nu$ conjecture is also satisfied by the critical behaviors of
the square-lattice Ashkin-Teller (AT) model. In the AT model one
considers two interacting sets of Ising spin variables,
$\sigma_{1,{\bm x}}$ and $\sigma_{2,{\bm x}}$, with Hamiltonian
\begin{eqnarray}
  H_{\rm AT} &=& - J \sum_{\langle {\bm x}{\bm y}\rangle} 
   \left( \sigma_{1,{\bm x}}  \sigma_{1,{\bm y}} 
  + \sigma_{2,{\bm x}}  \sigma_{2,{\bm y}} \right)    \label{ATmodel}\\
&& - K \,\sum_{\langle {\bm x}{\bm y} \rangle}
  \sigma_{1,{\bm x}} \sigma_{1,{\bm y}} \sigma_{2,{\bm x}}  \sigma_{2,{\bm y}},
\nonumber
\end{eqnarray}
where the sum extends over the bonds $\langle {\bm x}{\bm y} \rangle$
of a square lattice.  Note that the AT model can be formally
associated with the 2D two-component LGW theory 
with potential (\ref{cubicpot}).  The 2D AT model presents a
critical line $[J_c(K),K]$, from $K=-\infty$ to $K={1\over 4}\ln 3$
(where the 2D AT model can be mapped onto the 2D $q=4$ Potts model)
with ${\rm sinh}(J_c) = \exp(-2K)$. Along the line the exponent $\eta$
is constant, $\eta=1/4$, while the exponent $\nu$ varies continuously,
as~\cite{Baxter-book,Baxter-71,Baxter-72,WL-74,DR-79}
\begin{eqnarray}
  y_r(K) &=& {1\over \nu(K)} = {4w-3\pi \over 2(w-\pi)},\label{ytlAT}\\
  \cos w &=& e^{2K} \sinh(2K).\nonumber
\end{eqnarray}
Therefore, along the critical line $y_r$ monotonically increases from
$y_r=1/2$ for $K=-\infty$ to $y_r=3/2$ for $K={1\over 4}\ln 3$.
Ccorrespondingly, $\Sigma$ monotonically decreases from $\Sigma=5/4$ to
$\Sigma=1/4$, remaining always positive. It is interesting to observe
that in the AT model the conjecture is rigorously valid, as a
consequence of the results of Ref.~\cite{CS-00}, that proves $\gamma
\ge 1$ for the susceptibility.

Exact results have also been obtained for the O($N$) $\sigma$ model
for $N\in[-2,2]$~\cite{CH-80,Nienhuis-84,DF-84,CPRV-96}.  Defining the
parameter $z$ through the relation
\begin{equation}
  N = - 2 \cos(2\pi/z),\quad 1\le z \le 2,\quad -2\le N\le 2,
  \label{narel}
\end{equation}
then
\begin{eqnarray}
y_r = 4 - 2 z, \qquad \eta = {-z^2 + 4 z -3\over 2z},
\label{nl2exp}
\end{eqnarray}
implying
\begin{eqnarray}
\Sigma = 2 - \eta-y_r = {5 z^2 - 8z +3\over 2 z},
\end{eqnarray}
which is always positive for $z\ge 1$: it varies from zero
for $N=-2$ ($z=1$)  to 
7/4 for $N=2$ ($z=2$).

\section{Analytic results in the large-$N$ limit}
\label{lnvarious}

In this section, we present some analytic results obtained for large
values of $N$, which confirm the validity of the conjectured
inequality (\ref{nubound}).

\subsection{O($N$)-vector models}
\label{lnonvec}

We first consider LGW theories with $N$-component
fields ${\bm \varphi}({\bm x})$ and O($N$) global symmetry, which are
obtained by taking
\begin{equation}
u_{abcd} = {u\over 3}\,(\delta_{ab}\delta_{cd} +
\delta_{ac}\delta_{bd} + \delta_{ad}\delta_{bc})
\label{oncou}
\end{equation}
in Eq.~(\ref{genH}).

The large-$N$ expansions of the critical exponents 
for any $2<d<4$ are reported in 
Refs.~\cite{ZJ-book,MZ-03} and references therein.  
Setting
\begin{equation}
  X(d) = {4 \, \Gamma(d-2)\over \Gamma(d/2) \Gamma(2-d/2)
    \Gamma(d/2-1)^2},
  \label{wdef}
\end{equation}
to order $1/N$ we have
\begin{eqnarray}
  \eta &=& X(d) {4-d\over d N} + O(N^{-2}),
  \label{lnetaond}\\
  y_r &=&  d-2 + 
  X(d){2(d-1)(d-2) \over d N} + O(N^{-2}),\qquad
  \label{lnytond}
\end{eqnarray}
which implies 
\begin{eqnarray}
  \Sigma =  4-d - 
X(d) {2 d^2 - 7 d + 8  \over d N} + O(N^{-2}).         
  \label{lnWond}
\end{eqnarray}
The quantity $\Sigma$ is always positive for $2\le d < 4$, 
varying from $\Sigma=0$ for $d=4$ to
$\Sigma=2$ for $d=2$. In three dimensions, 
using $X(d=3)=8/\pi^2$, we obtain
\begin{eqnarray}
%  \eta &=& {8\over 3 \pi^2 N} - {512 \over 27 \pi^4 N^2} + O(N^{-3}),
%  \label{lnetaon}\\
%  y_r &=&
%  1 + {32\over 3 \pi^2 N} + {32(27\pi^2-16) \over 27 \pi^4 N^2} + O(N^{-3}),
%  \label{lnyton}\\
  \Sigma = 1 - {40\over 3 \pi^2 N} + O(N^{-2}).
  \label{lnWond3}
\end{eqnarray}

\subsection{O($M$)$\otimes$O($N$) LGW theories}
\label{lnonm}

We now focus on the $d$-dimensional O($M$)$\otimes$O($N$) LGW $\Phi^4$
theory defined by the Hamiltonian density
\begin{eqnarray}
{\cal H}&=& \sum_{a,c} (\partial_\mu \Phi_{ac})^2 + r \sum_{a,c}
  \Phi_{ac}^2  + u \Bigl( \sum_{a,c} \Phi_{ac}^2
\Bigr)^2 \nonumber
\\ &+& v \sum_{a,b,c,d}
      \Bigl(\Phi_{ac}\Phi_{bc} \Phi_{ad}\Phi_{bd} - \Phi_{ac}^2
        \Phi_{bd}^2\Bigr) \label{omn}
\end{eqnarray}
where $\Phi_{ac}$ is an $M\times N$ real matrix ($a=1,...,M$ and
$c=1,...,N$).  The symmetry group of the model is O($M$)$\otimes$ O($N$).

\begin{figure}[tbp]
  \includegraphics[width=0.6\columnwidth, clip]{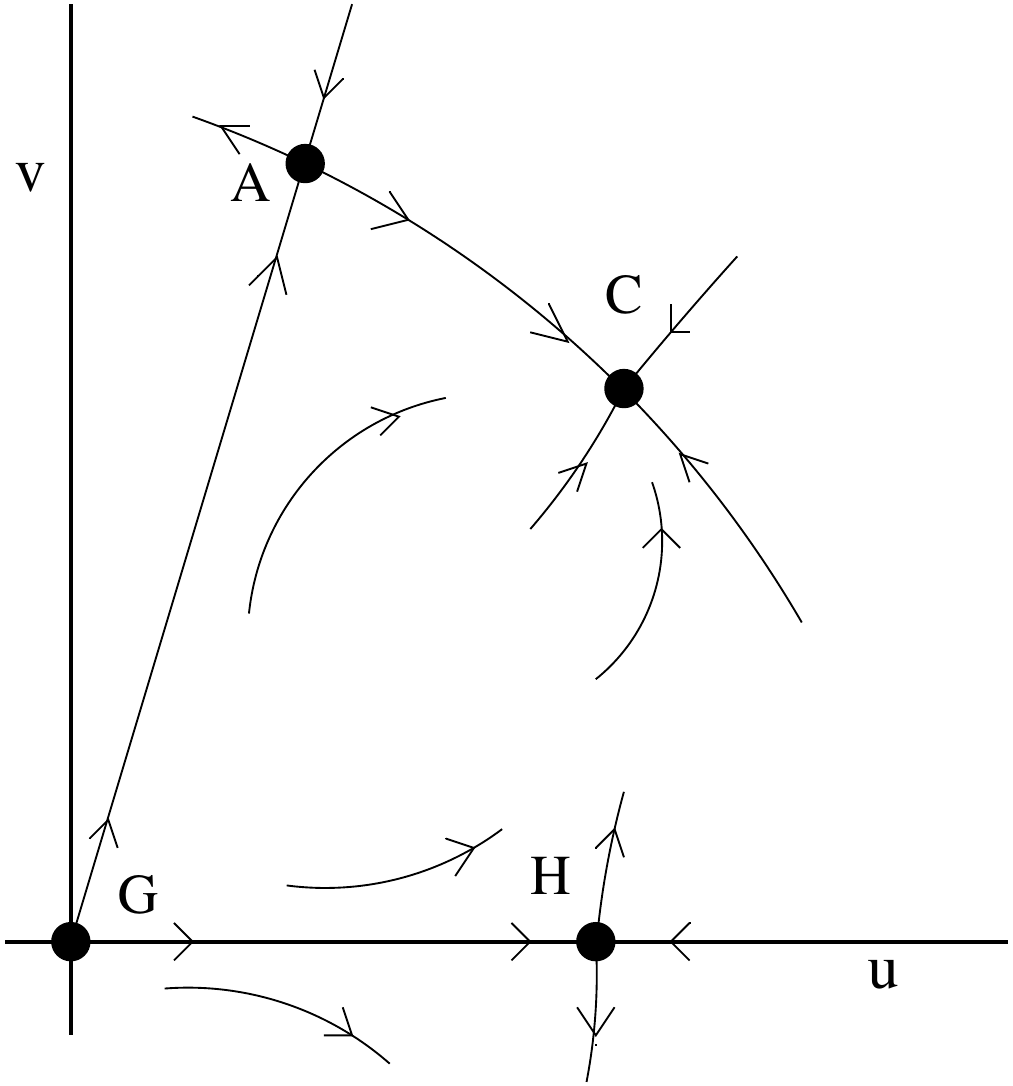}
\caption{ RG flow and FPs of the O($M$)$\otimes$O($N$) model in the
  large-$N$ limit. Tere are four FPs: the unstable Gaussian (G) and
  O($N$) symmetric (H) FPs, the stable chiral (C) FP and the unstable
  antichiral (A) FP.  }
\label{omnflow}
\end{figure}

The O($M$)$\otimes$O($N$) $\Phi^4$ model can be solved in the
large-$N$ limit for any fixed $M$ \cite{Kawamura-91,PRV-01b}.  For any
$M\ge 2$, there are four FPs: the Gaussian FP, the Heisenberg
O($M\times N$) FP, and two new FPs which we call chiral (C) and
antichiral (A). Fig.~\ref{omnflow} shows a sketch of the RG flow in
the quartic-coupling space.  In the large-$N$ limit, for any $M\ge 2$
and for any $2<d<4$, the stable FP is the chiral one, and all other
FPs are unstable.

It is useful to define the $d$-dimensional constants
\begin{eqnarray}
  &&A(d) = -{4\Gamma(d-2)\over \Gamma(2-d/2)\Gamma(d/2-1)\Gamma(d/2-2)
  \Gamma(d/2+1)},\nonumber\\
  &&B(d) = {(d-1)(d-2) \over (4-d)} A(d),
    \label{abdef} \\
  &&Y(d) =  A(d) + 2 B(d). \nonumber 
 \end{eqnarray}
At the stable chiral FP we have
\begin{eqnarray}
\eta_c &=& A(d) {M+1\over 2 N} + O(N^{-2}),
  \label{lnetaOMNc}\\
%  \nu_c &=&  {1\over d-2} - B(d){M+1\over (d-2)^2N} + O(N^{-2}),
%  \label{lnnuOMNc}\\
    y_{t,c} &=&  d-2 + B(d){M+1\over N} + O(N^{-2}),
\label{lnytOMNc}\\
\Sigma_c &=& 4-d - Y(d) {M+1 \over 2N} + O(N^{-2}).\qquad
\label{lnWOMNc}
\end{eqnarray}
At the unstable antichiral FP we have
\begin{eqnarray}
\eta_{ac} &=& A(d) {(M+2)(M-1)\over 2 M N} + O(N^{-2}),
  \label{lnetaOMNac}\\
%  \nu_{ac} &=&  {1\over 2} + B(d){(M+2)(M-1)\over 4MN} + O(N^{-2}),
%  \label{lnnuOMNac}\\
    y_{t,ac} &=&  2 - B(d){(M+2)(M-1)\over MN} + O(N^{-2}),
\label{lnytOMNac}\\
\Sigma_{ac} &=& Y(d) {(M+2)(M-1)\over 2MN} + O(N^{-2}).\qquad
\label{lnWOMNac}
\end{eqnarray}
One can easily check that $\Sigma_c$ and $\Sigma_{ac}$ are both
positive for any $2<d<4$ and any $M\ge 2$.  In particular, for $d=3$
we have 
\begin{eqnarray}
  &&\Sigma_c = 1 - {  20(M+1) \over 3\pi^2 N} + O(N^{-2}), \label{3dreslnomn}\\
  &&\Sigma_{ac} = {40 (M+2)(M - 1)\over 3 \pi^2 M N} + O(N^{-2}).
\nonumber
\end{eqnarray}

\section{Results for three-dimensional transitions}
\label{3dex}

We now verify that the $\nu$ conjecture holds at all known 3D
continuous transitions that are described by LGW theories.

\subsection{3D O($N$)-vector models}
\label{onsymm3d}

Accurate estimates of the critical exponents for the 3D O($N$) vector
models have been obtained by several studies.  For example, results
for $N=1$, 2, and 3 are reported in
Refs.~\cite{GZ-98,CPRV-02,Hasenbusch-10,KPSV-16,KP-17,FXL-18,
  Hasenbusch-21},
Refs.~\cite{GZ-98,CHPV-06,KP-17,Hasenbusch-19,CLLPSSV-20,Hasenbusch-25},
and
Refs.~\cite{GZ-98,CHPRV-02,HV-11,KP-17,Hasenbusch-20,Chester-etal-21},
respectively;  Refs.~\cite{Clisby-10,Clisby-17} report estimates for
the 3D self-avoiding walk (SAW) universality class ($N=0$);
Ref.~\cite{Hasenbusch-22} for $N\ge 4$.  The most accurate estimates
of $\nu$ and $\eta$ are reported in Table~\ref{3Dexp}.  In all cases,
$\Sigma=2-\eta-y_r$ is positive as a consequence of the smallness of
$\eta$ and of the fact that $\nu$ is significantly larger than 1/2.

\begin{table}
\caption{We report estimates of the critical exponents $\nu$
  and $\eta$, and of the difference
  $\Sigma=\Delta_\varepsilon-2\Delta_\varphi$, for some 3D continuous
  transitions.}
\label{3Dexp}
\begin{tabular}{clll} \hline\hline
 Model & $\quad \nu\quad$ & $\quad\eta\quad$ &$\quad \Sigma\quad$
  \\\hline

Gaussian & 1/2 & 0 & 0 \\
  
  Ising~\cite{KPSV-16} & 0.629971(4) & 0.03698(2) & 0.37565(1) \\

  O(2), XY~\cite{Hasenbusch-25} & 0.67172(2) & 0.03816(2) & 0.47312(5) \\

  O(3), H~\cite{Hasenbusch-20} & 0.71164(10) & 0.03784(5) & 0.5570(2) \\

  O($N\to\infty$) & $1 - {32\over 3 \pi^2 N}$ & ${8\over 3 \pi^2 N}$ & $1
  - {40\over 3 \pi^2 N}$ \\

  SAW [O($N\to 0$)]~\cite{Clisby-10,Clisby-17} &
  0.587597(7) & 0.03104(2) & 0.26711(4)\\

  Cubic $N=4$~\cite{Hasenbusch-23} & 0.7202(7) & 0.0371(2) &  0.574(1) \\

  Cubic $N\to\infty$ & 0.707902(6) & 0.03698(2) &
  0.54981(1) \\

  Cubic $N\to 0$, RDIs~\cite{HPPV-07} & 0.683(2) & 0.036(1) & 0.500(4) \\

  Percolation~\cite{XWLD-14} & 0.8762(12) & $-$0.0458(2) & 0.905(2)
  \\
  
\hline\hline  
\end{tabular}
\end{table}

\subsection{LGW theories with cubic anisotropy}
\label{cubic3d}

\begin{figure*}[htbp]
 \vspace{-3truecm} \hspace{-3truecm}
  \includegraphics[width=2.0\columnwidth, clip]{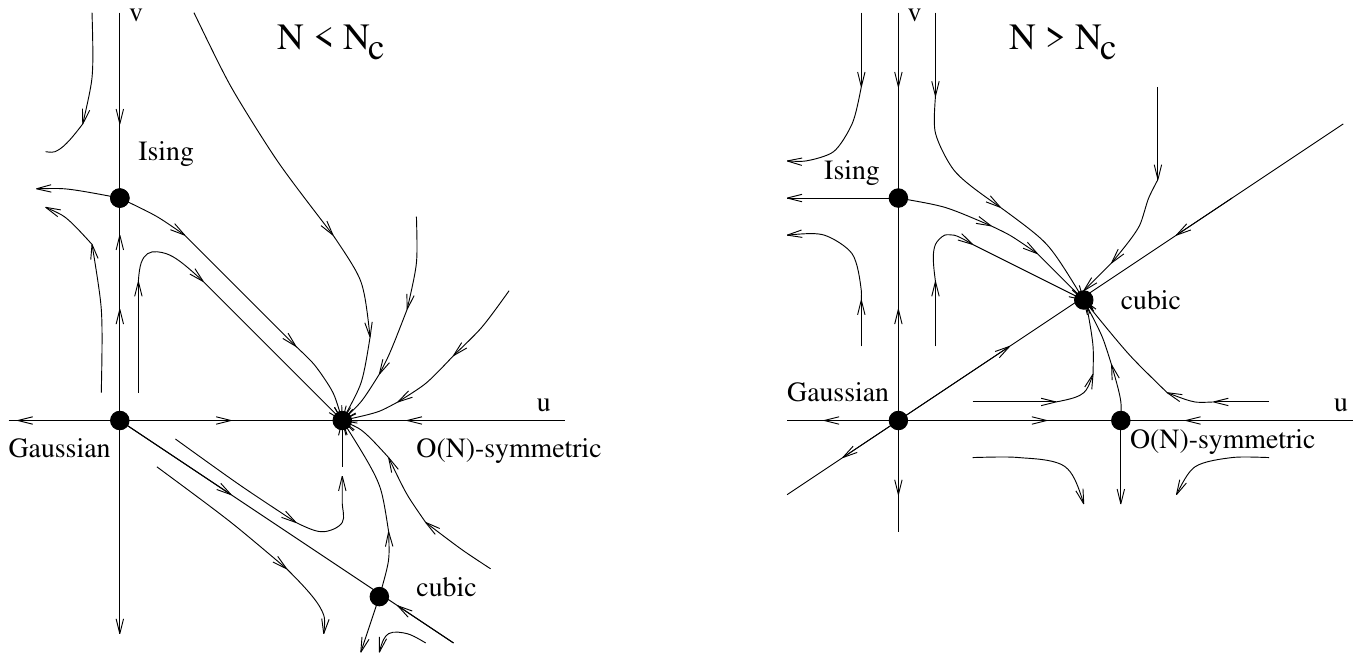}
 \vspace{-3.5truecm}
  \caption{The three-dimensional RG flow of the cubic model for $N=2$
    (left) and $N\ge 3$ (right), because $N_c\approx 2.9$ in three
    dimensions~\cite{Aharony-76,CPV-00,PV-02,HV-11,Chester-etal-21,
      Hasenbusch-23,Hasenbusch-24}.}
\label{cubicflow}
\end{figure*}

Another interesting class of continuous transitions are described by
LGW theories with cubic anisotropy.  The Hamiltonian density
is~\cite{Aharony-76}
\begin{eqnarray}
  {\cal H} =
  \sum_a [(\partial_\mu \varphi_a)^2 +  r \varphi_a^{2}] 
   + u (\sum_a \varphi_a^2)^2  +  v \sum_a \varphi_a^4 ,\quad
\label{cubicth}
\end{eqnarray}
where $\varphi_a$ is an $N$-component real field. The analysis of the
RG flow of the model shows the appearance of a cubic FP with
$u^*\not=0$ and $v^*\not=0$ beside the FPs present in O($N$) models.
The qualititative features of the RG flow as well as the stability of
the cubic FP depend on the number of components, see
Fig.~\ref{cubicflow}.  For $N<N_c$, the O($N$)-symmetric FP is stable
and the cubic one is unstable, while the opposite occurs for
$N>N_c$. In three dimensions, $N_c\approx 2.9$, see, e.g.,
Refs.~\cite{Aharony-74,HL-74,
  Aharony-76,CPV-00,PV-02,HV-11,Chester-etal-21,
  Hasenbusch-23,Hasenbusch-24} and references therein.

The presently known estimates of the critical exponents for the cubic
FP confirm $\Sigma > 0$. Indeed, for $N=3$ the cubic critical
exponents are very close to the Heisenberg exponents
\cite{Hasenbusch-23}: $\nu_{\rm cubic} - \nu_{\rm H} = -0.0006(1)$ and
$-0.00001 \lesssim \eta_{\rm cubic} - \eta_{\rm H} \lesssim 0.00007$,
where $\nu_{\rm H}$ and $\eta_{\rm H}$ are the critical exponents of
the O(3) universality class.  For $N=4$ results are reported in
Table~\ref{3Dexp}. For $N\to \infty$ one
obtains~\cite{Aharony-76,Aharony-73,Emery-75}
\begin{eqnarray}
  \eta = \eta_{\rm Is}+O\left( {1/N}\right),\quad
  \nu = 
       {\nu_{\rm Is}\over 1-\alpha_{\rm Is}}+O\left( {1/N}\right),
\label{largen}
\end{eqnarray}
where $\eta_{\rm Is}$, $\nu_{\rm Is}$, and $\alpha_{\rm Is}=2 -
3\nu_{\rm Is}$ are the critical exponents of the Ising model.  It
follows $\Sigma \approx 0.55 > 0$.

The model (\ref{cubicth}) is also interesting in the formal limit
$N\to 0$. In this case it describes the univeral features of
randomly diluted Ising (RDIs) systems~\cite{GL-76,PV-02}.  The
critical exponents (see Table~\ref{3Dexp} and
Refs.~\cite{PV-00,HPPV-07,PV-02} for a more complete list of results)
again satisfy $\Sigma>0$.

\subsection{O(2)$\otimes$O($N$) LGW theories}
\label{onm3d}

We finally consider the 3D O(2)$\otimes$O($N$) LGW theories with
Hamiltonian density~(\ref{omn}) for small values of $N$.  For $N=2$
and 3 these models may describe transitions in noncollinear frustrated
magnets and the normal-to-planar superfluid transition in $^3$He, see,
e.g., Refs.~\cite{Kawamura-98,PV-02,V-07} and references therein.  The
existence of continuous transitions and of corresponding 
stable RG FPs in these models is still
controversial. While the analyses of high-order perturbative
expansions and some numerical simulations favor the existence of
stable FPs associated with chiral universality classes for
$N=2,3$~\cite{Kawamura-98,PRV-01a,PV-02,CPPV-04,DPV-04}, other
theoretical studies do not confirm their existence, or at least their
stability~\cite{DMT-04,NT-14,RRSR-25}.  Note that these FPs for
$N=2,\,3$ are not related to  the FPs present close to four
dimensions for larger values of $N$, a phenomenon that also occurs in
the one-component Abelian Higgs theory~\cite{BPV-25-rev}, see also
App.~\ref{lnAH}.

We now verify the $\nu$ conjecture at the FPs found in
Refs.~\cite{PRV-01a,PV-02,CPPV-04,DPV-04} for the chiral
O(2)$\otimes$O(2) and O(2)$\otimes$O(3) transitions and for the $^3$He
superfluid transition (the last one has the same symmetry but a
different symmetry breaking pattern with respect to the chiral
transition), and verify that the corresponding estimates of the
critical exponents satisfy the $\nu$ conjecture. A field-theoretical
perturbative analysis gives $\nu\approx 0.6$ and $\eta\approx 0.09$
for the O(2)$\otimes$O(2) and O(2)$\otimes$O(3) chiral
FPs~\cite{PV-02,CPPV-04}, and $\nu = 0.60(5)$ and $\eta= 0.08(1)$ for
the $^3$He superfluid universality class ~\cite{DPV-04}.  In all cases
$\Sigma\approx 0.3 > 0$.

Other more complex LGW theories with multiparameter quartic terms have
also been considered in the literature. We mention here the theories
relevant for spin-density waves in cuprates and for the
finite-temperature chiral hadronic transition (such as
U($N$)$\otimes$U($N$)-symmetric LGW theories).  Field-theoretical
analyses have identified the relevant FPs and provided estimates of
the corresponding critical exponents \cite{DPV-06,PSV-08,PV-13}, which
satisfy $\Sigma>0$ in all cases.

\subsection{Percolation}
\label{percres}

The percolation problem can be related to the limit $q\to 1$ of the
Potts model, and also with a particular limit of the $\Phi^3$ field
theory, for which the upper critical dimension is $d=6$.  Therefore,
the $\nu$ conjecture is expected to hold as well.  This is confirmed by
results for 3D percolation~\cite{XWLD-14,BFMMPR-99} reported in
Table~\ref{3Dexp}.  For $d> 3$ the results of
Refs.~\cite{AMAH-90,Gracey-15,LW-19,Ziff-20} give $\Sigma\approx 0.65$
for $d=4$, $\Sigma\approx 0.32$ for $d=5$, confirming $\Sigma\ge 0$ in
all cases. Moreover, close to the upper critical dimension ($d=6$) one
can perform a $6-d$ expansion, using the appropriate $\Phi^3$ field
theory.  At one loop~\cite{Gracey-15,SVZ-20} one obtains
\begin{eqnarray}
&y_r= 2 - {5\over 21} \epsilon_6 + O(\epsilon_6^2),\quad
&\eta = -{1\over 21} \epsilon_6 + O(\epsilon_6^2),\nonumber\\
&\gamma = 1 + {2\over 21}\epsilon_6 + O(\epsilon_6^2),\quad
  &\Sigma = {2\over 7} \epsilon_6 + O(\epsilon_6^2),
\label{6mdexp} 
\end{eqnarray}
where $\epsilon_6 \equiv 6-d$.

\section{The Gross-Neveu-Yukawa fermionic model}
\label{lnGN3}

We now consider models with scalar fields coupled with fermionic ones,
showing that the $\nu$ conjecture also holds in these field
theories.  We focus on the GNY model defined by the Hamiltonian
density (\ref{lagrangianGNY}).  As discussed in
Ref.~\cite{ZJ-book,MZ-03}, the GNY theory can be related to the
Gross-Neveu model
\begin{equation}
{\cal H}_{\rm GN} = -\sum_{f=1}^{N_f} \bar{\psi}_f(\slashed{\partial}
+ m)\psi_f - \frac{g^2}{2N_f}\, \Big(\sum_{f=1}^{N_f} \bar{\psi}_f
\psi_f\Big)^2,
  \label{lagrangianGN}
\end{equation}
with attractive four-fermion interactions. The scalar field $\varphi$
appearing in the GNY model can be considered as an auxiliary real
scalar field associated with the bilinear fermionic operator $\sum_{f}
\bar{\psi}_f \psi_f$.  In model (\ref{lagrangianGNY}) the approach to
criticality is controlled by the quadratic term of the scalar field,
thus the operator $[\varphi^2]$ can be again identified as the
energy-density operator. Therefore we can check the validity of the
$\nu$-conjecture, and in particular of the inequality
$\Delta_\varepsilon\ge 2\Delta_\varphi$, also in this model.

We report the known leading terms of the large-$N_f$ expansion of the
exponents $y_r=1/\nu$ and $\eta_\varphi$ for the $d$-dimensional GNY
model~\cite{MZ-03}:
\begin{eqnarray}
  \eta_\varphi &=& 4-d - X(d) {(d-1)\over d N_f} + O(N_f^{-2}),
  \label{etaGNln}\\
  y_r &=& d-2 - X(d) {(d-2)(d-1)\over 2 d N_f} + O(N_f^{-2}),\quad
  \label{ytGNln}
\end{eqnarray}
where $X(d)$ is defined in Eq.~(\ref{wdef}). Therefore,
\begin{eqnarray}
  \Sigma = 2 - \eta_\phi - y_r =
  X(d) {d-1\over 2 N_f} +
  O(N_f^{-2}),\label{sigmaGNln}
  \end{eqnarray}
which is positive for any $2< d < 4$. 

An analogous analysis can be done for the chiral XY and Heisenberg GNY
model, using the large-$N_f$ results of
Refs.~\cite{Gracey-21,Gracey-18}.  For the Heisenberg GNY model we
have in three dimensions~\cite{Gracey-18}
\begin{eqnarray}  \eta_\varphi &=& 1 +
      {4(3\pi^2+16)\over 3\pi^4 N_f^2} +
      O(N_f^{-3}),\label{etaGNln3H}\\ y_r &=& 1 - {4\over \pi^2 N_f}
      + {9\pi^2+104\over 3\pi^4 N_f^2} +
      O(N_f^{-3}),\label{ytGNln3H}
\end{eqnarray}
thus
\begin{eqnarray}
  \Sigma = 2 - \eta_\varphi - y_r = {4\over \pi^2 N_f} +
  O(N_f^{-2}) > 0.\label{sigmaGNln3H}
\end{eqnarray}

Finally, we can verify the $\nu$ conjecture close to four dimensions. In
this limit the critical exponents of the GNY theory with Lagrangian
(\ref{lagrangianGNY}) corresponding to the stable FP
are~\cite{ZMMHS-17,INMS-18}
\begin{eqnarray}
  \eta_\varphi &=&  \epsilon { 2 N_f\over 3+2N_f}, \label{GNexp}\\
  y_r &=& 2 - \epsilon {10 N_f + 3 + \sqrt{4 N^2_f + 132 N_f + 9}
      \over 6(3+2N_f)}.\nonumber
\end{eqnarray}
Thus, 
\begin{eqnarray}
  \Sigma = \epsilon {3 - 2 N_f + \sqrt{4 N_f^2 + 132 N_f + 9}
    \over 6 (3+2N_f)},\label{GNkappaexp}
\end{eqnarray}
which is positive for any $N_f$.  Analogous results can be obtained
for the chiral XY and Heisenberg GNY models, using the results
reported in Refs.~\cite{ZMMHS-17,INMS-18}.

\section{The Abelian Higgs gauge theory}
\label{lnAH}

The SU($N$)-symmetric AH gauge field theory is obtained by minimally
coupling an $N$-component complex scalar field ${\bm \phi}({\bm x})$
with an electromagnetic U(1) gauge field $A_\mu({\bm x})$. The
Lagrangian density reads~\cite{HLM-74,ZJ-book}
\begin{eqnarray}
  &&{\cal L}_{\rm AH}={\cal L}_{\rm AH}({\bm \phi},{\bm A}) +
  {\cal L}_{\rm gf}({\bm A}),
  \label{AHFT}\\
&&{\cal L}_{\rm AH} = \frac{1}{4 g^2} \,F_{\mu\nu}^2+ |D_\mu{\bm\phi}|^2
+ r\, \bar{\bm \phi}\cdot{\bm \phi} + u \,(\bar{\bm \phi}\cdot{\bm
  \phi})^2,\qquad\nonumber\\
&&{\cal L}_{\rm gf}({\bm A}) = {1\over 2\zeta}
(\partial_\mu A_\mu)^2,
  \nonumber
\end{eqnarray}
where the gauge-fixing term ${\cal L}_{\rm gf}$ is necessary to obtain
a well-defined theory in perturbation theory. To go beyond
perturbation theory, one must consider a nonperturbative
regularization and show that the properly renormalized theory admits a
finite limit when the regularization is eliminated. If we use the
lattice regularization, this is possible only if the lattice model
admits a continuous transition with a diverging length
scale~\cite{BPV-25-rev}.

As in LGW theories, the exponent $\nu$ is related to the 
RG dimension $\Delta_\varepsilon=d-y_r$ of the 
quadratic scalar term $\varepsilon
\sim [\bar{\bm \phi}\cdot {\bm \phi}]$. The exponent $\eta_\phi$ 
is instead associated with the RG dimension $\Delta_\phi=(d-2+\eta_\phi)/2$
of a nonlocal gauge-invariant vector order-parameter or, in the gauge
fixed theory (\ref{AHFT}), of the Lagrangian field ${\bm \phi}({\bm x})$
in the hard Lorenz gauge $\zeta=0$~\cite{BPV-25-rev,BPV-24-ncAH}. 
We now compute $\Sigma = 2 - y_r - \eta_\phi$ and show that it is
positive at the stable FPs.

Close to four dimensions, a stable FP exists only for $N>N_*(4)=
90+24\sqrt{15} \approx 183$. At order $\epsilon$ the critical
exponents are~\cite{HLM-74,IZMHS-19}
\begin{eqnarray}
  \eta_\phi &=& - \epsilon {6 + 3\zeta\over N}, \label{AHexp}\\
  y_r &=& 2 - \epsilon {N^2 + N - 54 + (N+1)\sqrt{N^2 - 180 N - 540}
      \over 2 N (N+4)}.\nonumber
\end{eqnarray}
For $\zeta=0$  we obtain
\begin{equation}
  \Sigma = \epsilon {N^2 + 13 N - 6 + (N+1)\sqrt{N^2 - 180 N - 540}
    \over 2 N (N+4)},\label{AHkappaexp}
\end{equation}
which is positive for $N>N_*(4)=90+24\sqrt{15}$ (actually, $\Sigma$
turns out to be positive for any value of the gauge-fixing parameter
$\zeta$).

The difference $\Sigma=2-y_r-\eta_\phi$ of the 3D AH theory can be
also computed in the large-$N$ limit, using the large-$N$ results reported
in Refs.~\cite{HLM-74,IKK-96,KS-08},
\begin{eqnarray}
  \eta_\phi &=& - {20+8\zeta \over \pi^2 N} + O(N^{-2}),
  \label{lnetaAH}\\
  y_r &=& 1 + {48\over \pi^2 N} + O(N^{-2}),
\label{lnytAH}
\end{eqnarray}
see also Refs.~\cite{BPV-22,BPV-25-rev} for comparisons with numerical
Monte Carlo results. For $\zeta=0$ we obtain
\begin{eqnarray}
\Sigma = 1 - {28\over \pi^2 N} + O(N^{-2}). 
\label{lnWAH}
\end{eqnarray}

To obtain finite-$N$ estimates in three dimensions one should consider
the related lattice models, which admit several transition lines
belonging to different universality classes. We focus here on the
Coulomb-Higgs charged transitions that are controlled by the FPs of
the AH field theory. These transitions are continuous for $N\ge
N_*(3)$. Monte Carlo simulations indicate $N_*(3)=7(2)$~\cite{BPV-22}.
Consistent estimates are obtained from field-theoretical analyses,
$N_*=12(4)$~\cite{IZMHS-19}, and from an analysis of the R\'enyi
entanglement entropy, $N_*\approx 7$~\cite{SJCSJY-25}. The estimates
of $y_r$ and $\eta_\phi$ are close to the large-$N$ estimates and
satisfy the bound $\Sigma > 0$.

 For $N<N_*(3)$ the Coulomb-Higgs transitions are first order, with
 the notable exception of the one-component AH model (relevant for
 superconductivity, see,
 e.g. Refs.~\cite{HLM-74,DH-81,HT-96,Herbut-book}). In this case the
 model presents a stable FP that is not connected with the stable FP
 found in the $\epsilon$-expansion analysis. It belongs to the inverted XY
 universality class, related to the standard XY universality class by
 duality~\cite{NRR-03,BPV-24-decQ2} (more precisely, duality relates
 energy observables in the two models). Thus, the exponent $\nu$ is
 the same as in the XY model, $\nu=0.6717$. On the other hand, the
 exponent $\eta_\phi$ differs from that of the standard XY
 universality class, as the order parameter in the AH model is a
 nonlocal gauge-invariant operator which is unrelated with the field
 in the standard XY model~\cite{BPV-24-ncAH,BPV-25-rev}.  Numerically,
 one finds $\eta_\phi=-0.74(4)$, so $\Sigma=2-\eta_\phi-y_r = 1.25(4)
 > 0$.

\end{document}